\documentclass{PoS}

\title{QCD and low-x physics at a Large Hadron electron Collider}

\ShortTitle{QCD and low-x physics at a Large Hadron electron Collider}

\author{\speaker{Paul Laycock}\\
         \thanks{on behalf of the LHeC Study Group.}\\
        University of Liverpool\\
        E-mail: \email{laycock@hep.ph.liv.ac.uk}}

      \abstract {The Large Hadron electron Collider (LHeC) is a
        proposed facility which will exploit the new world of energy
        and intensity offered by the LHC for electron-proton
        scattering, through the addition of a new electron
        accelerator. This contribution, which is derived from the
        draft CERN-ECFA-NuPECC Conceptual Design report (due for
        release in 2011), addresses the expected impact of the LHeC
        precision and extended kinematic range for low Bjorken-x and
        diffractive physics, and detailed simulation studies and
        prospects for high precision QCD and electroweak
        fits. Numerous observables which are sensitive to the expected
        low-x saturation of the parton densities are explored. These
        include the inclusive electron-proton scattering cross section
        and the related structure functions $F_2$ and $F_L$, as well as
        exclusive processes such as deeply-virtual Compton scattering
        and quasi-elastic heavy vector meson production and
        diffractive virtual photon dissociation. With a hundred times
        the luminosity that was achieved at HERA, salient expectations
        for the LHeC include the complete determination of all light
        and heavy quark parton distributions for the first time, the
        high precision extraction of the gluon density, the
        determination of the strong coupling constant to per-mil
        accuracy and the precision study of the running of the
        electroweak mixing angle.}

\FullConference{The 2011 Europhysics Conference on High Energy Physics-HEP 2011,\\
                July 21-27, 2011\\
                Grenoble, Rh\^one-Alpes France}

\begin{document}

\section{Introduction}

The Large Hadron electron Collider (LHeC) concept is to collide a new
high energy electron beam with an existing LHC proton or ion beam,
while simultaneously running the full LHC physics programme.  The
Conceptual Design Report is currently under review from CERN-appointed
referees.  Only a selection of highlights are presented here, the
reader is strongly encouraged to find further information on the LHeC
web site~\cite{LHeC-URL}.

\section{Physics Highlights}

Among the highlights of the LHeC is its impact on our understanding of
proton structure\footnote{It should be noted that, thanks to the
  electron-ion colliding mode of operation, the improvement in our
  knowledge of nucleon structure coming from the LHeC can be compared
  to the impact of HERA on proton structure.}.  The expected size of
the final dataset is of order $100 \,{\rm fb^{-1}}$, some two orders
of magnitude more than HERA.  A challenging but feasible large
detector acceptance down to $1^{\circ}$ gives a kinematic extension of
a factor of $\approx 20$ in $x$ and $Q^2$ with respect to the HERA
experiments.  The implications for proton structure measurements are
correspondingly large.  For example, the uncertainty on the gluon PDF,
shown in Figure 1, would be greatly improved, reaching percent level
precision across the full kinematic range.

Similar levels of improvement are expected in the valence quark
distributions, and also on the strange quark content of the proton
thanks to high precision charged current measurements.  The
improvement in kinematic coverage and precision of heavy flavour
measurements, courtesy of silicon tracking based on existing
technologies, would allow much progress in our understanding of heavy
flavour quarks in the proton.  The precision of the data will be
sensitive to different theoretical approaches to the treatment of
heavy flavours in fits and allow the massive and massless approaches
to be distinguished.  Such fits would be able to constrain the strong
coupling constant to the per-mil level of precision.


\begin{figure}
  \begin{center}
    \includegraphics[width=0.9\columnwidth]{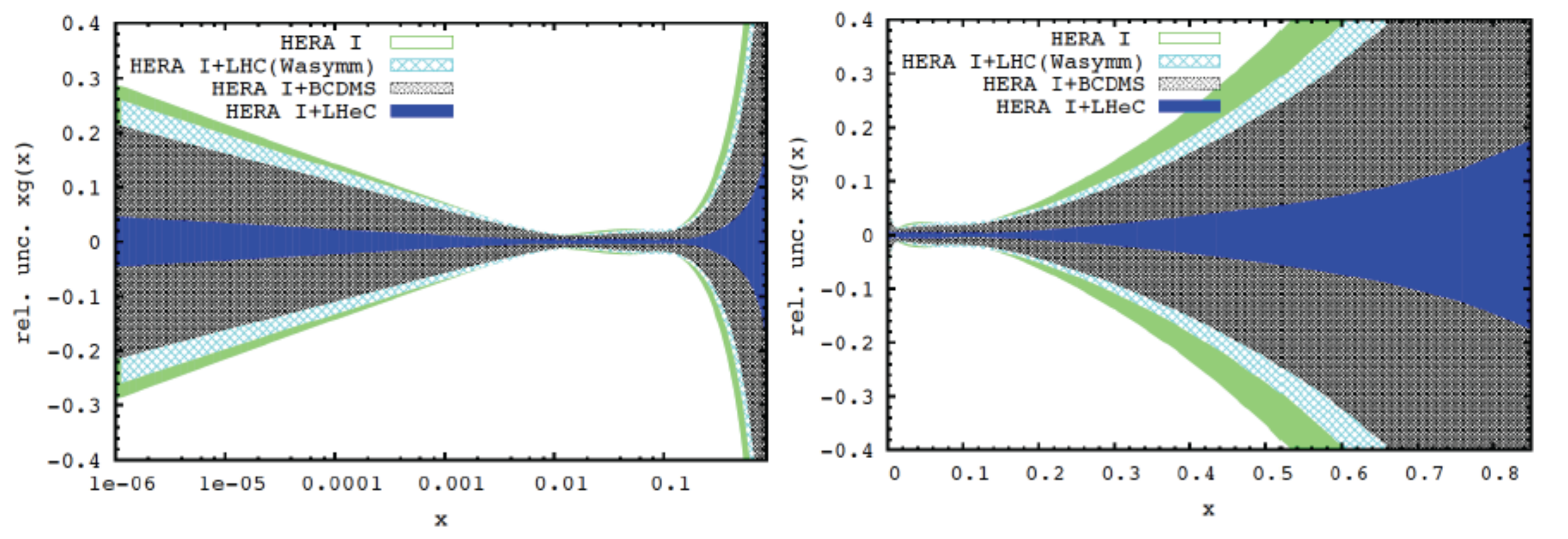}
    \caption{The relative uncertainty on the gluon on a linear and log
      $x$ scale, in scenarios with and without the LHeC.}\label{Fig:g}
  \end{center}
\end{figure}

\subsection{Parton Saturation}

The rise of the gluon PDF at low $x$ cannot continue forever without
violating unitarity, but conclusive proof of saturation has yet to be
seen.  Thanks to its kinematic coverage, the LHeC should be able to
resolve this.  Figure 2 shows longitudinal structure function
pseudo-data generated using a model which includes saturation
(AAMS09), compared to a fit to the same data which does not allow for
saturation (NNPDF Fit).  The fit clearly fails to describe the data.

\begin{figure}
  \begin{center}
    \includegraphics[width=0.6\columnwidth]{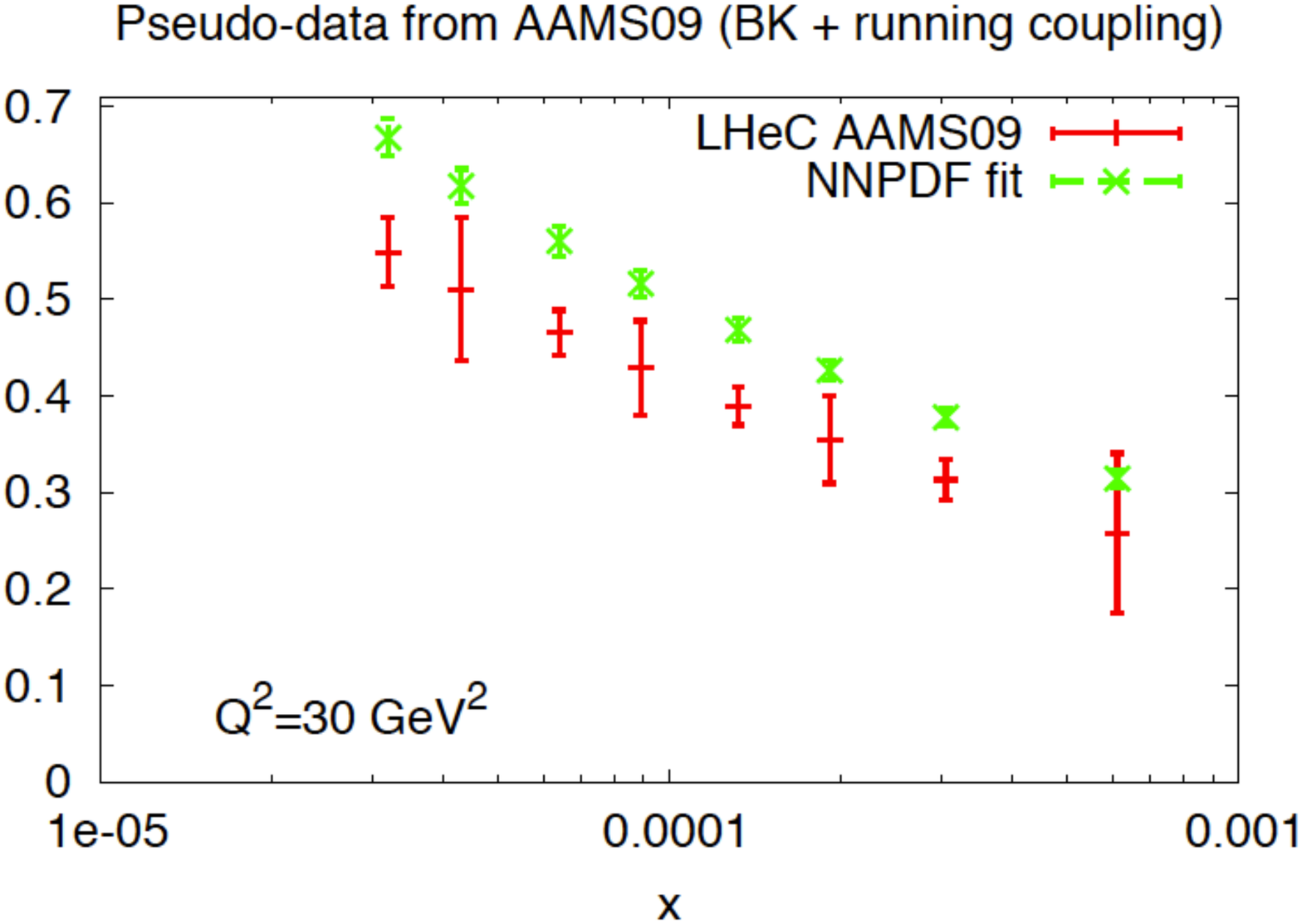}
    \caption{The longitudinal structure function $F_L$ from
      pseudo-data which includes saturation (AAMS09) compared to a fit
      to the same data which does not allow for
      saturation.}\label{Fig:sat}
  \end{center}
\end{figure}

\subsection{Diffraction}

A related phenomenon is diffraction, where colour-singlet
configurations dominate.  Exploiting the rapidity gap toplogy, the
experimental precision will again be improved, while the larger centre
of mass energy means that larger-mass diffractive final states can be
accessed than at HERA.  This opens up the possibility to measure
diffractively produced $W$ and $Z$ bosons, where the $W$ in particular
will test the quark flavour symmetry assumptions normally made when
modelling diffraction.  Exclusive diffractive vector meson production,
especially the $J/\Psi$, has long been used as a test of perturbative
QCD and to look for saturation effects.  The large extension in the
kinematic range may mean that exclusive $J/\Psi$ data will be able to
distinguish between models of saturation.

\section{Conclusion}

The LHeC would provide a new window on QCD and low $x$ physics,
providing data with a kinematic range and precision that will
greatly improve our knowledge of both.

\end{document}